\documentclass[11pt]{article}

\usepackage{comment}

\usepackage{amssymb}
\usepackage{amsmath}
\usepackage{amsthm}
\usepackage{color}

\usepackage{fullpage}

\title{On Arroyo-Figueroa's Proof that P $\neq \np$}

\author{Mandar Juvekar, David E. Narv\'aez\thanks{Supported in part by NSF grant CCF-2030859 to the Computing Research Association for the CIFellows Project.}, and Melissa Welsh\thanks{Supported in part by NSF grant CCF-2006496.} \\ Department of Computer Science \\ University of Rochester \\ Rochester, NY 14627, USA
}

\date{March 28, 2021}

\newcommand{\cond}{\,\mid \:}

\newcommand{\pe}{\mbox{\rm P}}

\newcommand{\np}{\mbox{\rm NP}}

\mathchardef\mhyphen="2D
  \newtheorem{theorem}{Theorem}[section]

  \newtheorem{definition}[theorem]{Definition}

\begin{document}
\maketitle

\begin{abstract}
We critique Javier Arroyo-Figueroa's paper titled ``The existence of the Tau one-way functions class as a proof that $\pe \neq \np$,'' which claims to prove $\pe \neq \np$ by showing the existence of a class of one-way functions. We summarize our best interpretation of Arroyo-Figueroa's argument, and show why it fails to prove the existence of one-way functions. Hence, we show that Arroyo-Figueroa fails to prove $\pe \neq \np$.
\end{abstract}

\section{Introduction}\label{s:intro}

The question of whether the complexity classes $\pe$ and $\np$ are equal is one of the most well-known open problems in theoretical computer science, a resolution to which would have far-reaching consequences in computer science, mathematics, and beyond. In the paper titled ``The existence of the Tau one-way functions class as a proof that $\pe \neq \np$,'' Arroyo-Figueroa~\cite{arroyo-fig:p-neq-np} claims to have answered this question in the negative: that $\pe$ is in fact \textit{not} equal to $\np$.

Arroyo-Figueroa's argument is based on the notion of one-way functions: functions that are easy to compute but hard to invert. Formally, one-way functions are defined as follows.

\begin{definition} \label{def:one-way}
    A function $f: \{0,1\}^* \to \{0,1\}^*$ is a one-way function if it can be computed with a polynomial-time algorithm, but any polynomial-time probabilistic algorithm $F$ that attempts to compute the inverse of $f$ succeeds with negligible probability. That is, for large enough inputs $x$ and all positive integers $c$,
    \begin{equation*}
        \mathrm{Pr}(f(F(f(x), \mathrm{unary}(n))) = f(x)) < n^{-c},
    \end{equation*}
    where $n$ is the size of $x$ and $\mathrm{unary}(n)$ is a bit string of $n$ ones.
\end{definition}

The existence of one-way functions has long been an open problem. It is known that if one-way functions exist then $\pe \neq \np$ (see \cite[Proposition 1]{sel:j:one-way}\footnote{It is worth mentioning that the one-way functions studied in Selman's work~\cite{sel:j:one-way} are worst-case (complexity-theoretic) one-way functions, whereas Arroyo-Figueroa considers average-case (cryptographic) one-way functions.
However, this distinction does not matter in our context because the existence of average-case one-way functions implies that of worst-case one-way functions. Thus if Arroyo-Figueroa's proof of the existence of average-case one-way function were to hold, it would imply $\pe \neq \np$.}), which means constructing a one-way function would resolve the $\pe$ vs.\ $\np$ dispute. Arroyo-Figueroa~\cite{arroyo-fig:p-neq-np} claims to have constructed a class $T$ of functions which are one-way, thus showing that $\pe \neq \np$.

In this short paper we go over Arroyo-Figueroa's argument and describe why it fails to show that the functions constructed are actually one-way, and is hence insufficient as a proof that $\pe \neq \np$.\footnote{Our critique is based on the most recent version of Arroyo-Figueroa's paper available at the time of this writing. Namely, we are critiquing Version 4 of arXiv.org report 1604.03758, last revised on October 17, 2016.}

\section{Arroyo-Figueroa's Argument}\label{s:recap}
Arroyo-Figueroa begins by constructing a class of functions $T$, where each $\tau \in T$ maps a bit sequence $x$ of length $n$ to another bit sequence $y$ of length $n$.
The class $T$ relies on $M$, a set of $n$ uniformly-distributed random bit matrices, each unique and of size $n \times n$, and $H$, a set of independent universal hash functions, as defined below.
\begin{definition} \label{def:hash-fam}
   A family of functions $H = \{h: U \to [m]\}$ is a universal family if for all $x, y \in U$ with $x \neq y$,
    \begin{equation*}
        \underset{h\in H}{\mathrm{Pr}} (h(x) = h(y)) \leq 1/m.
    \end{equation*}
    In other words, if a hash function $h$ is drawn randomly from a universal family $H$, then the probability of two keys in $U$ colliding is at most $1/m$. 
\end{definition}
 
In Arroyo-Figueroa's work, $U = \{0,1\}^n$ for some positive integer $n$, and $m < 2^n$.
For each $\tau \in T$ in Arroyo-Figueroa's construction and input $x \in \{0,1\}^n$, $\tau(x)$ is computed by repeatedly appending bits to an initially empty string $y$. The bits appended are computed based on traversals of the matrices $M_i \in M$. These traversals are decided using $h(x)$ for various $h \in H$. The execution time of the algorithm is bounded by $O(n^2)$, meaning $\tau(x)$ can be computed in time polynomial in $n$.
 
 Arroyo-Figueora then claims that every polynomial-time randomized algorithm attempting to find the inverse of any $\tau$ given $y$ succeeds with negligible probability. If this claim holds, then Arroyo-Figueora has successfully proven that all members of $T$ are in fact one-way functions. It would then necessarily follow that $\pe \neq \np$. Arroyo-Figueroa walks through a series of steps to calculate upper bounds on the probability of finding at random an $x$ that, when input into the algorithm previously described, produces a given output $y$. Arroyo-Figueroa defines and assumes the following:
 \begin{enumerate}
     \item The event $F$ is defined as finding at random one member of the preimage $\tau^{\mbox{-}}(\{y\})$. Here, and throughout the paper, the preimage of a set $S \subseteq B$ under a function $f : A \to B$ refers to the set $f^-(S) = \{a \in A \cond f(a) \in S\}$.
     \item The event $F_i$ is defined as finding at random one element from the preimage $\tau^{\mbox{-}}(\{y\})$ ``with a path'' to $y_i$. While the exact meaning of this terminology isn't specified in Arroyo-Figueroa's paper, our interpretation is that an ``element of the preimage with a path to $y_i$'' is an element $a$ from the domain of $\tau$ such that the $i$th bit of $\tau(a)$ is $y_i$. Hence $F_i$ is the event of finding at random a string that leads to the correct $i$th bit, which means $\mathrm{Pr}(F) = \mathrm{Pr}(\bigcap_{i \in [n]} F_i)$ as observed by Arroyo-Figueroa \cite[Lemma~10]{arroyo-fig:p-neq-np}.
     \item The size of the domain of each $h_i \in H$ is $2^n$.
     \item The size of the preimage of $h_i$ is given by $\lVert h_i^{\mbox{-}}(\{y_i\}) \rVert=2^n/8$ by definition of the universal family $h_i$ is drawn from.
     \item All $h_i \in H$ are independent.
     \item $F_i$ and $F_j$ are not independent for $i \neq j$.
 \end{enumerate}
The proof begins by bounding $\mathrm{Pr}(F_i)$, then $\mathrm{Pr}(F_i|F_j)$, and then the probability of the intersection $\mathrm{Pr}(\bigcap_{i \in [n]} F_i) = \mathrm{Pr}(F)$, which leads to the conclusion that $\mathrm{Pr}(F)<n^{-c}$ for sufficiently large $n$ and all positive integers $c$. Referring back to the original definition of a one-way function (see Definition~\ref{def:one-way}), Arroyo-Figueora claims this proves that all members $\tau$ of $T$ are one-way functions, and therefore that $\pe \neq \np$.
 
\section{Critique}\label{s:critique}

\subsection{\boldmath The Construction of $T$} \label{subseq:inconsist}

An understanding of $T$'s accuracy is hindered by inconsistencies and ambiguities in the class's construction. Though it is possible these ambiguities have no effect on the functions' one-wayness, these limitations are worth mentioning for the purpose of future research that may extend Arroyo-Figueroa's work.

Firstly, each function $\tau \in T$ is said to map a bit sequence of length $n$ to another bit sequence of length $n$. However, for every bit in the input sequence, the algorithm appends $n$ bits to the output sequence (see \cite[Page 5]{arroyo-fig:p-neq-np}). This means that $T$ would instead need to be defined as $\{\tau \cond \tau:\{0,1\}^n\to\{0,1\}^{n^2}\}$. The most dangerous implications of this inconsistency would be incorrectly calculating the size of hash functions' preimages and the probability of randomly selecting inputs that map to the correct outputs.

Secondly, the set of hash functions $H$ and their definitions are ambiguous. All hash functions in $H$ are defined as being drawn at random from either $\{ h :\{0,1\}^n\to\{0,1\}^{\log(n)}\}$ or $\{h : \{0,1\}^n \to \{0,1\}^3\}$. However, the definition of universal families of hash functions used (see Definition~\ref{def:hash-fam}) defines hash functions as mapping bit strings to integers, and not other bit strings. Furthermore, multiple steps of the given algorithm add $1$ to the output of a hash function. The result of this is meant to represent coordinates of a matrix. Nevertheless, the paper does not specify how a string of bits is meant to be added with and converted to an integer within the bounds of $n$. One could assume that Arroyo-Figueora intended these functions to hash inputs into $\log(n)$ or $n$ different buckets. In this case, the hash functions would instead be defined as coming from families $\{h : \{0,1\}^n \to \{0,1,\ldots,\log(n)\}\}$ and $\{h : \{0,1\}^n \to \{0,1,2,3\}\}$, which would be consistent with Definition~\ref{def:hash-fam}. However, it could also be assumed that the bit sequence is meant to be treated as a binary number and translated to its decimal form. This ambiguity could generate unintended coordinates, which could in turn produce unintended results.

\subsection{\boldmath The One-Wayness of Functions in $T$}

Looking at the proof of the one-way nature of $T$, we observe a deeper issue with the proof strategy. As is evident from Definition~\ref{def:one-way}, to prove that a function is one-way, one must demonstrate that it is computable in polymonial time, and that \emph{every} probabilistic polynomial-time algorithm that attempts to invert it succeeds with only a negligible probability. Arroyo-Figueroa properly shows the former, but fails to establish the latter. Arroyo-Figueroa attempts to prove the latter claim in Section~9 of~\cite{arroyo-fig:p-neq-np}. This section contains various computations and arrives at the main theorem of the paper. The theorem is reproduced below.

\begin{theorem}[{\cite[Theorem 1]{arroyo-fig:p-neq-np}}]
    Let $y = \tau(x)$ for some $x \in \{0,1\}^n$. Let $F$ be the event of finding at random one member of the preimage $\tau^-(\{y\})$. The probability of $F$ is bounded by
    \begin{equation*}
        \mathrm{Pr}(F) < n^{-c},
    \end{equation*}
    where $c$ is any positive integer.
    \label{t:fail}
\end{theorem}

From this theorem, Arroyo-Figueroa concludes that $\tau$ is a one-way function (see \cite[Corollary 3]{arroyo-fig:p-neq-np}). However, the theorem as stated is insufficient to prove this corollary. In particular, the event $F$ in the theorem statement that is shown to occur with probability less than $n^{-c}$ is defined as the ``event of finding \emph{at random} one member of the preimage $\tau^-(\{y\})$.'' Thus what the theorem shows is that if one were to pick randomly from $\{0,1\}^n$, the odds of picking an element in the preimage $\tau^-(\{y\})$ are small. But this doesn't necessarily entail that \emph{every algorithm} would have such a low probability of success. Indeed, most probabilistic algorithms in the literature employ techniques more sophisticated than picking at random.

There are many polynomial-time functions for which picking at random to find the inverse yields a low probability of success, but whose inverses are in fact computable in polynomial time. One such example is the function $f: \{0,1\}^n \to \{0,1\}^n$ that inverts each bit of its input (the bitwise `not' operator). Clearly, $f$ can be computed and inverted in polynomial time, which means $f$ is not one-way. Now consider, given $y \in \{0,1\}^n$, trying to find an element of the preimage $f^-(\{y\})$ by picking at random. Since $f$ is bijective, the preimage contains exactly one element. The domain of $f$, $\{0,1\}^n$, contains $2^n$ elements. So the probability of picking an element of the preimage at random is $2^{-n}$, which is asymptotically less than $n^{-c}$ for any positive integer $c$. In other words, $f$ satisfies the conclusion of Theorem~\ref{t:fail}. Yet, $f$ definitely does not satisfy the corollary that Arroyo-Figueroa draws from the conclusion of Theorem~\ref{t:fail}.

It is worthwhile to ask whether this is indeed the intended interpretation of Theorem~\ref{t:fail}. After all, ``finding at random'' as used in the theorem could also be interpreted as picking using some procedure that involves randomness, which is much closer to what probabilistic algorithms in the literature do. This is likely the construction Arroyo-Figueroa intended, as is evident from the concluding paragraph of the paper in question, which says that ``[i]t was also proved that any random algorithm that attempts to find the inverse of any function in $T$ has negligible probability of success''~\cite[Section 10]{arroyo-fig:p-neq-np}. However, the methodology of the proofs of Theorem~\ref{t:fail} and the lemmas leading up to it support our interpretation. Throughout Section~9, Arroyo-Figueroa computes probabilities of events by dividing the number of favorable outcomes by the size of the sample space. For instance, the proof of Lemma~6 contains the claim that for each $h_i$ in a particular subset of $H$, the probability of finding an $x$ in the preimage $h_i^-(\{y_i\})$ is equal to $1/8$. This claim is justified in the proof by saying that the size of the preimage is $2^n/8$ and the size of $\{0,1\}^n$ is $2^n$, and so the probability is $(2^n/8)/2^n = 1/8$. The same method of computing probabilities appears in other lemmas used in the proof of Theorem~\ref{t:fail}. This method computes the probability of picking elements from the sample space uniformly at random, but does not extend to arbitrary randomized procedures. Therefore, we believe that our interpretation of Theorem~\ref{t:fail}, while probably different from that intended by Arroyo-Figueroa, is the one actually proved in the paper, assuming the steps in the proofs of the individual lemmas are correct and the inconsistencies in Section~\ref{subseq:inconsist} are fixed.

\section{Conclusions}

Arroyo-Figueroa's argument attempts to construct a family $T$ of functions that are one-way. If shown to be one-way, these functions would be evidence that $\pe \neq \np$. Unfortunately the argument as it stands is flawed. The construction of $T$ is unclear and possibly not coherent since the functions in $T$ do not have the claimed type (namely $\{0,1\}^n \to \{0,1\}^n$). Even if these inconsistencies are fixed and we assume---without deciding here---that the proofs of the lemmas and the theorem in the paper are correct, Arroyo-Figueroa fails to show that the probability that any probabilistic polynomial-time algorithm can find an inverse of a function $\tau \in T$ is negligible, as is required by the definition of one-wayness. We thus believe that Arroyo-Figueroa's argument does not establish $\pe \neq \np$.

\section*{Acknowledgments}

We thank Michael C.\ Chavrimootoo, Lane A.\ Hemaspaandra, and Arian Nadjimzadah for their helpful comments and suggestions on earlier drafts of this critique. Any remaining errors are the responsibility of the authors.

\bibliography{bibliography}

\begin{thebibliography}{AF16}

\bibitem[AF16]{arroyo-fig:p-neq-np}
Javier~A. Arroyo-Figueroa.
\newblock The existence of the {T}au one-way functions class as a proof that
  {P} != {NP}.
\newblock Technical Report arXiv:1604.03758~[cs.CC], Computing Research
  Repository, \mbox{arXiv.org/corr/}, October 2016.

\bibitem[Sel92]{sel:j:one-way}
A.~Selman.
\newblock A survey of one-way functions in complexity theory.
\newblock {\em Mathematical Systems Theory}, 25(3):203--221, 1992.

\end{thebibliography}

\end{document}